# Symmetry and polarity of the voltage-controlled magnetic anisotropy studied by the Anomalous Hall effect

*Vadym Zayets, Takayuki Nozaki, Hidekazu Saito, Akio Fukushima, Shinji Yuasa*
Spintronics Research Center,
National Institute of Advanced Industrial Science and Technology (AIST), Tsukuba, Japan

Abstract:
*The voltage-controlled magnetic anisotropy (VCMA) effect in FeB and FeB/W films was measured by four independent methods. All measurements are consistent and show the same tendency. The coercive field, Hall angle, anisotropy field, the magnetization switching time and retention time linearly decrease when the gate voltage increases and they linearly increase when the gate voltage decreases.*

The VCMA effect describes the fact that in a capacitor, in which one of electrodes is made of a thin ferromagnetic metal, the magnetic properties of the ferromagnetic metal are changed, when a voltage is applied to the capacitor. For example, under an applied voltage the magnetization direction of the ferromagnetic metal may be changed[1–3] or even reversed[4,5]. This magnetization-switching mechanism can be used as a data recording method. When an electrical pulse reverses magnetization direction, the data is memorized in the ferromagnetic metal by means of its two opposite magnetization directions. Such a recording mechanism is fast and energy-efficient. It may be used in the magnetic random access memory (MRAM)[6] and the all-metal transistor[7].

The first demonstration of the VCMA effect was in a FePt film immersed into a liquid electrolyte[2]. The change of the coercive field under a gate voltage was detected. The demonstration of the VCMA in an all-solid device[1] and the demonstration of a high-speed magnetization reversal by the VCMA[4] opened a possibility of the fabrication of a new type of high-speed low- power- consumption MRAM and triggered an intense interest in this topic.

Several plausible physical mechanisms of the VCMA effect has been suggested. It is understood that the gate voltage affects the interfacial perpendicular magnetic anisotropy (PMA) for the following reason. Inside a metal the electrical field is screened by the conduction electrons and cannot penetrate deep inside the metal. As a result, the voltage, which is applied to the capacitor dielectric (gate), may penetrate into and affect only the few uppermost atomic layers of the metal near the gate. However, the change of magnetic properties of the uppermost layer by the gate voltage affects the magnetic properties of the whole film and the change by the gate voltage may be substantial. The strength of the interfacial PMA depends on the local magnetization, atom arrangement and bonding at the interface[8]. The VCMA may occur only because some of these parameters are affected by the gate voltage. The voltage-controlled change of the PMA may occur due to the spin-dependent screening[9], the electric-field induced dipole formation[10], the accumulation/depletion of conduction electrons at the interface[11–13], the Rashba effect[14,15] and the voltage modulation of the magnetic dipole[16]. The features of these mechanisms are a fast response and a long endurance. Additionally, the voltage-controlled change of the PMA may occur due to the voltage-induced redox reaction[17–19], the electromigration[20] and the piezoelectric effect[21]. However, the latter mechanisms have a slower response speed and a lower endurance.

There are several methods to measure the VCMA effect. In the first method, the voltage dependency of the anisotropic field is measured. The anisotropic field may be measured using the Anomalous Hall effect (AHE)[22,23] or the tunnel magneto-resistance (TMR)[24,25]. In both cases, the magnetization direction is measured as a function of the in-plane magnetic field, from which the value of the anisotropic field is evaluated. The half of the product of the anisotropic field and the saturation magnetization equals to the PMA energy[8]. Therefore, this method allows estimation of the voltage dependence of the PMA energy as well. The different polarities of the voltage dependence of the PMA energy were reported. A linear voltage dependency of the PMA energy with a negative slope was measured in Ta/FeCoB[26,27], Cr/Fe[25], Au/Fe[1], Ru/Co$_2$FeAl[28] and with a positive slope in Ru/FeCo[29]. The negative slope means that the PMA energy increases linearly when the gate voltage decreases and consequently the number of conduction electrons at the interface between the ferromagnetic metal and the gate dielectric decreases. A symmetric dependence vs the gate-voltage polarity was measured in Cr/Fe[25,30] and Cr/Fe/Cr[31].

In the second method to evaluate the VCMA effect, the voltage dependency of the coercive field is measured. The TMR, Hall and magneto-optical Kerr measurements are used. A linear voltage dependency of the coercive field with a negative slope was measured in Ta/FeCoB[23,32], Au/FeCo[3], Ru/Co$_2$FeAl[28] and with a positive slope in Pd/FePd[33], Ta/FeCoB[34], Pt/Co[22,35]. The third method to measure the VCMA is the voltage-induced ferromagnetic resonance excitation[36–38]. The explanation of the resonance VCMA effect is more complex and will not be discussed here.

In this work, we have studied the VCMA effect in a Fe$_{0.8}$B$_{0.2}$ film and a Fe$_{0.8}$B$_{0.2}$/W multilayer using 4 independent measurements. For each sample, in addition to the conventional measurements of the voltage-dependence of the anisotropic and coercive fields, the voltage-dependence of the Hall angle, the magnetization–switching time and retention time were measured and analyzed. The polarity and symmetry of the different possible contributions into the VCMA effect were examined and their correlation with the experimental data are discussed. The PMA of a thicker FeB/W multilayer is comparable to that of a thinner FeB film[39]. Except the thickness, the materials, structure and properties of these films are almost similar. It allows us to compare the VCMA effect in a thicker and a thinner film.

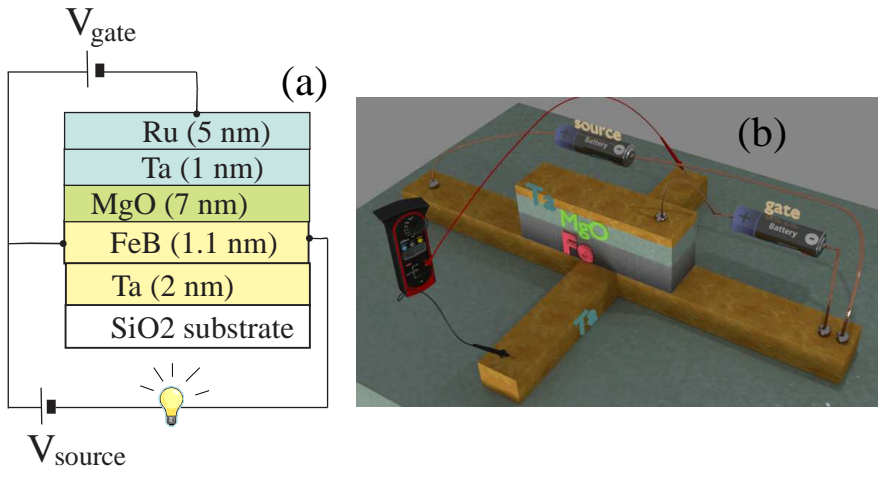

Figure 1. (a) Layer stack of FeB film (b) top-view of measurement setup.

The VCMA was measured using the Anomalous Hall effect (AHE)[22,23]. For VCMA measurements, the AHE configuration has several advantages compared to the MTJ configuration. Firstly, there is no undesirable influence of the dipole magnetic field from the reference electrode on measured VCMA properties and there is no undesirable influence of the spin transfer torque due to the flow of the spin-polarized current from the reference electrode. Secondly, different materials of the gate electrode can be tested. The MTJ configuration is limited to a specific ferromagnetic metal, which has to provide a sufficient TMR. There is no such limitation for the Hall configuration. Thirdly, in the AHE configuration the voltage-dependence of several magnetic parameters can be measured in a single sample (e.g. 4 independent parameters in this study). Each measurement reveals different features of the VCMA effect.

The samples were fabricated on a Si/SiO2 substrate by sputtering. Figure 1(a) shows a stack of layers. A Ta (2 nm) was used as a buffer layer and a Ta (1 nm)/Ru(5 nm) was used as a gate electrode. A FeB (1.1 nm) or a FeB(0.8 nm)/W(1.5 nm)/FeB(0.8 nm) multilayer was used as a ferromagnetic layer. There is a week ferromagnetic exchange coupling between two FeB layers[39]. A thick MgO (7 nm) layer was used to suppress the tunneling current. A nanowire of different width between 100 and 3000 nm with a Hall probe (Fig.1(b)) was fabricated by the argon milling. The positive gate voltage means that a positive voltage was applied to the non-magnetic electrode. In order to increase the break-down voltage and to suppress the oxygen diffusion in the gate the following growth procedure was used. At first, a 1 nm MgO was deposited at room temperature, next the sample was annealed at $220^0$ C for 30 minutes and the remaining of the MgO gate oxide was grown at $220^0$ C. Three 30-minute growth interruptions after each 1 nm of growth were used to improve the MgO crystal quality.

The change of the coercive field under the gate voltage is not large and requires a high measurement precision. The estimated precision of our measurements was about 0.9 Oe. A repeated measurement after a 1-month interval is well fit within the precision. The coercive field $H_C$ is the magnetic field, at which the magnetization direction is reversed along its easy axis. The magnetization switching is a thermally-activated process, which is described by the Néel model[40–42]. In the Néel model it is assumed that the switching between two stable magnetization states occurs when the energy of a thermal fluctuation becomes larger than the energy barrier $E_{barrier}$ between states. The switching time $t_{switching}$ is described by the Arrhenius law[42] as

$$t_{switching} = f^{-1} \cdot e^{\frac{E_{barrier}}{kT}} = f^{-1} \cdot e^{\frac{E_{PMA}}{kT}\left(1-\frac{H}{H_{anis}}\right)^2} \sim t_{retention} \cdot e^{-\frac{H \cdot M_{eff}}{kT}} \quad (1)$$

where $f$ is attempt frequency; $E_{barrier} = E_{PMA}\left(1-\frac{H}{H_{anis}}\right)^2$ is the energy barrier, which separates two opposite magnetization states[41]; $E_{PMA} = 0.5 \cdot H_{anis} \cdot M_{eff}$ is the PMA energy; $H_{anis}$ is the anisotropy field; $M_{eff}$ is the effective magnetization; and $t_{retention}$ is the retention time, which is the average time of magnetization reversal without any external magnetic field. Since it was measured $\frac{H_C}{H_{anis}} \approx 2-5\%$ for all our samples and the magnetic field was scanned in the vicinity of $H_c$ ($H \approx H_C$), the condition $\frac{H}{H_{anis}} \ll 1$ was used to simplify Eq.(1). From Eq.(1), the coercive field $H_C$ can be calculated as

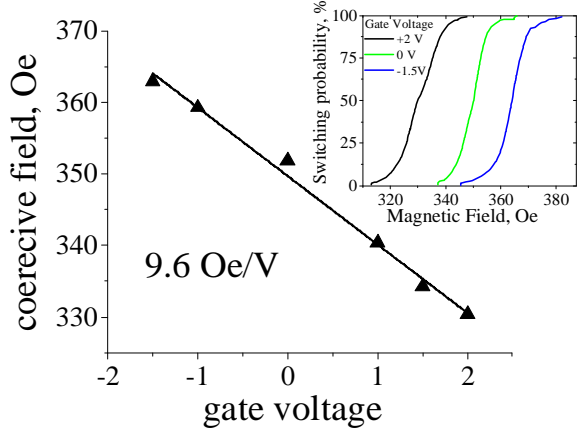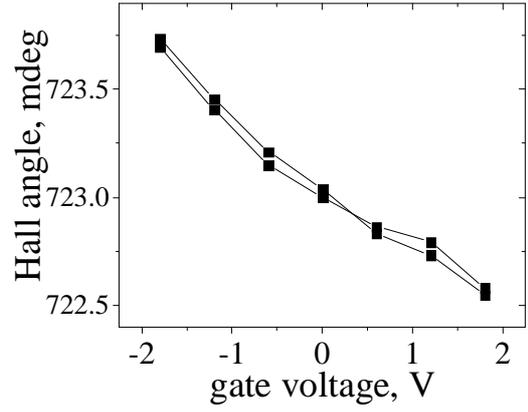

Figure 2 Coercive field vs voltage measured in FeB/W multilayer. Inset shows switching probability from magnetization-up to magnetization-down direction as function of the magnetic field

Figure 3. Hall angle as function of the gate voltage in FeB film. the gate voltage was scanned from a negative to a positive value and back to the negative value

$$H_C = \frac{kT}{M_{eff}} \ln\left(\frac{t_{switching}}{t_{retention}}\right) \quad (2)$$

From Eq.(2), the $H_C$ depends on the switching time $t_{switching}$ or in other words on the measurement time. The longer the measurement time is, at the smaller value of magnetic field the magnetization is switched. For example, when measurement time equals to $t_{retention}$, the $H_C$ equals to zero. In this paper we refer to all values of $H_C$ for the measurement time of one second. Due to the dependence of $H_C$ on the measurement time, the required precise measurements of $H_C$ can only be achieved either by method of a sweep magnetic field[43] or by method of a pulsed magnetic field[44].

A substantial number of measurements[45] and statistical analysis are required in order to obtain the $H_C$ with a required high precision. In order to shorten the measurement time, we have developed an optimized measurement method of $H_C$, which consists of two sets of measurements. In the first measurement, the magnetic pulses of the gradually-increased amplitude were applied and the magnetization switching field was measured. In the second measurement, the magnetic pulses of the constant amplitude were applied and the magnetization switching time was measured. The required measurement precision can be reached faster by combining the data from these two measurements. The used method is similar to that which was used in Ref.[44]. The details of the optimization of measurements will be described elsewhere.

Figure 2 shows measured $H_C$ of the FeB/W film as the function of the gate voltage. All measurement points fit well to a straight line. The inset of Fig.1 shows the magnetization switching probability at different gate voltages. The curves practically do not overlap. The change of the coercive field is substantial. It is 9.6 Oe/V. In the case of the FeB sample, the slope is smaller (3.7 Oe/V).

Figure 3 shows the gate-voltage dependency of the Hall angle[46] in the FeB film. All measurement points fit well in a straight line with a negative slope. There isn't any hysteresis loop, which is a feature of a sample of a lower crystal quality of the gate oxide. In the case of the AHE, the Hall angle is proportional to the magnetization of the ferromagnetic metal, the spin polarization of the conduction electrons and the strength of spin-orbit interaction[47]. Assuming that the latter two parameters are not affected by the gate voltage, the data indicates that the magnetization of the FeB and FeB/W films increases under a negative gate voltage and decreases under a positive gate voltage. The polarity of this dependence is opposite to that observed in Pt/Co/MgO [22]. There is a clear relation between the gate-voltage dependencies of the coercive field and the Hall angle (the magnetization).

Figure 4 shows the dependence of the anisotropy field $H_{anis}$ on the gate voltage for the FeB film. The anisotropy field is defined[8,26] as the in-plane magnetic field, at which initially-perpendicular magnetization turns completely into the in-plane direction. The inset of Fig.4 shows the measured in-plane component of the magnetization as a function of the in-plane magnetic field. The slope is substantially different at a different gate voltage. The overcrossing of each line with the x-axis gives the $H_{anis}$. From Fig. 4, the $H_{anis}$ linearly depends on the gate voltage. It increases at a negative gate voltage and decreases at a positive gate voltage. There is a clear relation between the gate-voltage dependencies of $H_c$ and $H_{anis}$. In samples, in which the voltage dependence of $H_c$ is larger, the voltage dependence of $H_{anis}$ is larger as well. The $E_{PMA}$ can be calculated from the $H_{anis}$ and the saturation magnetization[8,26]. For the FeB film the estimated change of the $E_{PMA}$ is 50 fJ/V m.

Figure 5 shows the dependence of the switching time on the magnitude of an external perpendicular magnetic field. On a logarithmic scale, the switching time is linearly proportional to the magnetic field. Therefore, the magnetization switching is well described by the Arrhenius law (Eq.1) even in the case when magnetic properties are changed by a gate voltage. The switching time becomes longer at a negative gate voltage and shorter at a positive gate voltage. From Eq.(1), the slope of the lines is proportional to $M_{eff}$ and the horizontal offset is proportional to $t_{retention}$ and consequently to $E_{PMA}$. As can be seen from Fig. 5, the slope for each line is nearly the same. It means that the $M_{eff}$ is not significantly affected by the gate voltage. The $M_{eff}$ describes the average bulk magnetization during the magnetization reversal. It is not sensitive to a small change of magnetization in a thin region near the interface. In contrast, the Hall angle (Fig.3) is more sensitive to such a small change[47]. It can be concluded from data of Figs. 3 and 5 that the absolute value of the magnetization is modulated by the gate voltage, but the magnetization change occurs in only a small region in close proximity to the interface.

In summary, the gate-voltage dependence of the magnetization-switching time, retention time, Hall angle, anisotropic and coercive fields have been studied in the FeB and FeB/W films. All four independent measurements of the VCMA effect are consistent and show the same tendency. The coercive and anisotropic fields, the Hall angle, the magnetization-switching time and retention time linearly depend on the gate voltage. They all increase under a negative gate voltage and decrease under a positive gate voltage. The linear voltage dependence of the anisotropic fields (Fig.4) and the retention time (Fig.5) proves that the PMA strength changes under a gate voltage. The measured relation between the voltage-dependence of the different magnetic properties of a ferromagnetic film may help to understand the physical origin of the VCMA effect.

The VCMA effect is found to be substantially larger in the FeB/W multilayer than in the FeB thin film. This demonstrates that the VCMA effect is not limited by an ultrathin film (thickness <1 nm) and the VCMA may be substantial even in a thicker film.

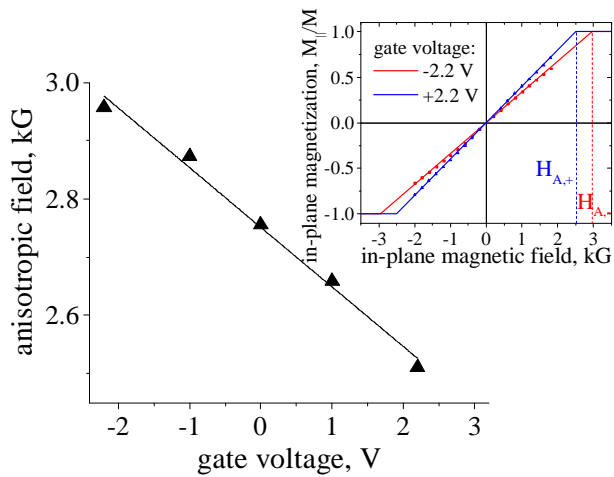

Figure 4. Anisotropic field vs gate voltage in FeB film. Inset shows in-plane magnetization as a function of the in-plane magnetic field. $H_{A+}$, $H_{A-}$, are anisotropic fields for $V_{gate}$= +2.2 V, -2.2 V respectively.

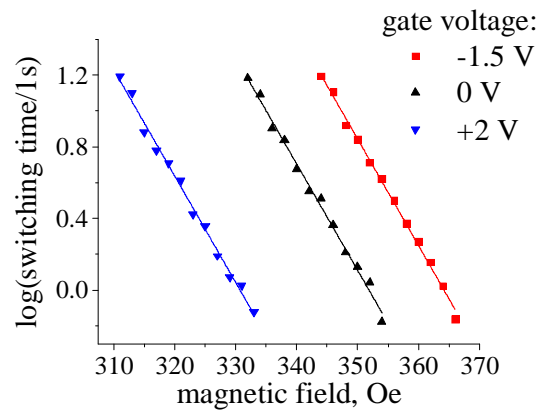

Figure 5. Magnetization switching time in FeB/W multilayer as a function of the perpendicular magnetic field. The line crossing with the x-axis (y=0) gives the coercive field of 363, 351, 330 Oe and the crossing with y-axis (x=0) gives retention time of $10^{21.66}$, $10^{20.88}$, $10^{19.64}$ seconds at $V_{gate}$ = -1.5, 0, +2 V, respectively.

Comment1: The Hall angle is linearly proportional to the Hall resistance. In contrast to the Hall resistance, the Hall angle[47] is only material dependent and does not depend on geometrical parameters like a film thickness.